\documentclass[twocolumn]{aastex63}
\usepackage{epsfig}
\usepackage{natbib}
\usepackage{amsmath}
\usepackage{hyperref}

\usepackage{xcolor}
\definecolor{forestgreen}{rgb}{0.13, 0.75, 0.13}

\newcommand{\p}{^\prime}

\newcommand{\OM}{\Omega_{\rm M}}

\newcommand{\partialmusys}{\frac{\partial\vec{\mu}}{\partial \rm{sys}}}

\newcommand{\improvement}{1.5}

\begin{document}

\title{Binning is Sinning (Supernova Version):  The Impact of Self-Calibration in Cosmological Analyses with Type Ia Supernovae}

\author{Dillon Brout}
\affil{Center for Astrophysics, Harvard \& Smithsonian, 60 Garden Street, Cambridge, MA 02138, USA}
\affil{NASA Einstein Fellow}
\email{djbrout@gmail.com}
\author{Samuel R. Hinton}
\affil{School of Mathematics and Physics, University of Queensland,  Brisbane, QLD 4072, Australia}
\author{Dan Scolnic}
\affil{Department of Physics, Duke University, Durham, NC, 27708, USA.}

\submitjournal{Astrophysical Journal Letters}

\begin{abstract}
Recent cosmological analyses (e.g., JLA, Pantheon) of Type Ia Supernova (SNIa) have propagated systematic uncertainties into a covariance matrix and either binned or smoothed the systematic vectors in redshift space. 
We demonstrate that systematic error budgets of these analyses can be improved by a factor of $\sim\improvement\times$ with the use of unbinned and unsmoothed covariance matrices. To understand this, we employ a separate approach that simultaneously fits for cosmological parameters and additional self-calibrating scale parameters that constrain the size of each systematic. We show that the covariance-matrix approach and scale-parameter approach yield equivalent results, implying that in both cases the data can self-calibrate certain systematic uncertainties, but that this ability is hindered when information is binned or smoothed in redshift space. 
We review the top systematic uncertainties in current analyses and find that the reduction of systematic uncertainties in the unbinned case depends on whether a systematic is consistent with varying the cosmological model and whether or not the systematic can be described by additional correlations between SN properties and luminosity. Furthermore, we show that the power of self-calibration increases with the size of the dataset, which presents a tremendous opportunity for upcoming analyses of photometrically classified samples, like those of Legacy Survey of Space and Time (LSST) and the Nancy Grace Roman Telescope (NGRST). However, to take advantage of self-calibration in large, photometrically-classified samples, we must first address the issue that binning is required in currently-used photometric methodologies.
\end{abstract}




\bigskip
\section{Introduction}

Reduction of systematic uncertainties in cosmological analyses with Type Ia Supernova (SNIa) is critical for future progress in constraining cosmological models with SNeIa.
This is because recent measurements have shown that systematic uncertainties in the measurement of the equation-of-state of dark energy $w$ are on the level of statistical uncertainties \citep{Betoule2014,Scolnic18,Brout18b,Jones19}.  To achieve their stated goals, the Legacy Survey of Space and Time (LSST; \citealp{Ivezic}) SN survey and the Nancy Grace Roman Space Telescope (NGRST; \citealp{Spergel15}) require significant reductions to the systematic floor \citep{Hounsell18,Mandelbaum18}, well below the floors of current state-of-the-art analyses. Notably, recent cosmological analyses with SNeIa have all used similar frameworks for calculating the impact of systematic uncertainties on measurements of cosmological parameters. In this paper, we show that those frameworks have unintentionally inflated the impact of systematic uncertainties and that future prospects for the systematic floor are expected to significantly improve as datasets grow in size.

The most widely used method to account for systematic uncertainties in SNIa analyses was introduced in \cite{Conley2010}.  There, a nominal set of distance moduli for the SN sample is computed, and for each systematic, distance moduli are recalculated and the differences between the two sets of distances are propagated into a covariance matrix.  The contributions from each source of systematic uncertainty are added to generate a full systematic covariance matrix that is used for cosmological-model constraints. In both \cite{Scolnic18} and \cite{Brout18b}, there were $>$70 individual sources of systematic uncertainty; the majority of which were calibration-related due to uncertainties in zeropoints and filter measurements. However, there are several non-calibration related systematics that contribute similarly as much to the overall error budget; these include uncertainties in the intrinsic scatter model \citep{Kessler13,Scolnic13}, the SALT2 model training \citep{Guy2010}, modeling of the color-luminosity relation \citep{bs20}, Milky Way extinction maps \citep{Schlafly16}, and more. While systematic uncertainties may depend on a variety of astrophysical, cosmological, or survey specific properties, recent analyses such as JLA \citep{Betoule2014}, Pantheon \citep{Scolnic18}, and DES3YR \citep{Brout18b} have diverged slightly from the \cite{Conley2010} approach when they began smoothing, binning, and marginalizing over much of this complex information. 

There were multiple motivations for smoothing or binning in past studies. In JLA, they `smoothed' the systematics in redshift space for three subsamples of their data: the entire low-$z$ subsample, the SDSS subsample and the SNLS subsample. The motivation for the smoothing was to prevent the contributions of additional statistical fluctuations due to the systematics. After JLA, the BEAMS with Bias Corrections (BBC: \citealt{BBC}) was developed to correct for biases in the recovered nuisance parameters and to account for both populations of SNeIa and non-Ia for utility in upcoming photometric analyses (DES, LSST, NGRST). In the BBC framework, the SN distances are binned in redshift space because the BEAMS likelihood requires a marginalization over Ia and core collapse populations in bins of SNe. While for spectroscopic analyses BBC has the capability to report unbinned data vectors, the spectroscopic analyses of Pantheon and DES3YR built their covariance matrices from the vectors of binned distance modulus differences for each systematic. An additional motivation for binning in BBC was to prepare for the increased computational intensity required in inverting matrices with samples of thousands of SNe.

To better understand the impact of binning on the systematic uncertainties in cosmological parameter constraints, we employ a separate method for estimating systematic uncertainties, introduced in \cite{Faccioli11}, hereafter F11.  While this method is similar to \cite{Conley2010} in that it first calculates a nominal set of distances and then computes a change in the distance modulus vector for each systematic, F11 do not compute a covariance matrix. Instead, F11 fit simultaneously for a scale parameter of the systematic magnitude alongside cosmological model parameters. F11 show in this method that the data can ‘self-calibrate’ the size of certain systematics when the systematic residuals and the cosmological model exhibit some amount of orthogonality. Here, we investigate the claim of F11 that this self-calibration ability is novel to their scaling method. 
 
In this work, we show using both the covariance matrix and scale parameter approaches that binning the data vector results in significantly poorer constraints on cosmological parameters in comparison to when the effects of systematics on individual SNe are preserved. In Section~\ref{sec:method}, we introduce the covariance-matrix approach and the systematic-scaling approach for computing cosmological constraints with systematic uncertainties. In Section~\ref{sec:toymodel}, we develop a toy model to demonstrate why binning or smoothing the data vector removes the ability for the SN dataset to perform self-calibration for many systematics in both approaches. In Section~\ref{sec:datasim}, we show improvements in systematic constraints on realistic survey simulations and a dataset of $\sim1500$ real SNeIa. Lastly, in Section~\ref{sec:discussion} we discuss the implications for analyses of photometric samples in the future and in Section~\ref{sec:conclusion} we give our conclusions.

\section{Methods}
\label{sec:method}
In this section, we present an overview of the determination of SNIa distance moduli and two separate methods for calculating the impact of systematic uncertainties in cosmological-parameter fitting.

\subsection{Distance Moduli and the Cosmological Model}

We follow the SALT2 framework \citep{Guy2010,Betoule2014} as implemented in \texttt{SNANA} \citep{Kessler2009} to fit SNIa light-curve parameters for standardization.  SALT2 fits of broadband photometric light-curves return four parameters: $m_B$, the log of the fitted light-curve amplitude $x_0$; $x_1$, called `stretch' which corresponds to light-curve width; $c$, the light-curve colour; and $t_0$, the date of peak brightness. Distance modulus values  ($\mu$) are inferred following the Tripp estimator \citep{Tripp98} such that
\begin{equation}
\label{eq:tripp}
    \mu = m_B + \alpha x_1 - \beta c - M + \delta_{\rm bias}
\end{equation}
where $M$ is the absolute magnitude of a SNIa with $x_1=c=0$, $\alpha$, $\beta$ are the correlation coefficients for $x_1$ and $c$ respectively and $\delta_{\rm bias}$ is a bias-correction term determined from simulations.

Distance uncertainties ($\sigma_\mu$) are computed from the uncertainties in the light-curve fit parameters and their covariance ($C$) and are given by
\begin{multline}
  \label{sigmastat}
   \sigma^2_\mu = C_{m_B,m_B} + \alpha^2 C_{x_1,x_1} + \beta^2 C_{c,c} + \\ 2\alpha C_{m_B,x_1} - 2\beta_{\rm SALT2} C_{m_B,c} - 2\alpha\beta C_{x_1,c} ~ .
\end{multline}

The redshifts and distance moduli are compared to a cosmological model.  For this analysis, we utilize a flat $w$CDM model with equation-of-state of dark energy $w$ and matter density $\Omega_M$ such that luminosity distances are be described by

\begin{eqnarray}
\label{eq:modeldist}
  d_L(z,w,\OM)  & = & (1+z) \frac{c}{H_0}  \int_0^z  \frac{ dz^{\prime}} { E(z^{\prime}) }~,\\
   E(z)  & = & \left[    \OM(1+z)^3  + (1-\OM)(1+z)^{3(1+w)}  \right]^{1/2}~.  \nonumber
\end{eqnarray}
Here, the Hubble constant ($H_0$) is included in the cosmological model, but is degenerate with $M$ from Eq.~\ref{eq:tripp}. 
Finally, we define a model distance modulus as
\begin{equation}
\label{eq:modeldist}
\mu_{{\rm mod}} = 5 \textrm{log}_{10}(d_L) + 25 ~. 
\end{equation}

\subsection{Binning}
In recent cosmological analyses \citep{Scolnic18,Brout18b,Jones19}, the BBC method is utilized to measure distance modulus values for the SNe. When analyzed on spectroscopically classified datasets, BBC outputs both binned and unbinned versions of the Hubble diagram. 
\cite{Scolnic18} and \cite{Brout18b} utilized the binned versions after verifying that for a spectroscopically confirmed sample, cosmological fits to the binned and unbinned Hubble diagrams without any systematic uncertainties yield equivalent constraints on cosmological parameters. 
 
The redshift-binned Hubble diagram is computed as:

\begin{equation}
        \label{eq:zz}
    \langle z \rangle_{\rm bin}   =  \langle z_i \in {\rm bin} \rangle~,
\end{equation}

\vspace{-.1in}

\begin{equation}
    \langle\mu\rangle_{\rm bin}   =  \langle \mu^i-\mu_{\rm fid}^i \in {\rm bin} \rangle + \mu_{\rm fid}(\langle z \rangle_{\rm bin})~,
        \label{eq:muz}
        \vspace{.05in}
\end{equation}
where $\langle\rangle$ express distance modulus uncertainty weighted averages and where in Eq.~\ref{eq:muz} a fiducial cosmology, $\mu_{\rm fid}(z) = \mu_{\rm mod}(z,w=-1,\Omega_M=0.3)$, is first subtracted off before averaging and then added back in for the binned average redshift $\mu_{\rm fid}(\langle z \rangle_{\rm bin})$. In the following sections, we assess systematic uncertainty quantification in both binned and unbinned approaches.

\subsection{Systematics with Covariance Matrices}
\label{sec:covmats}

Analyses such as \cite{Scolnic18} and \cite{Brout18b} adapted the method introduced by \cite{Conley2010} for the propagation of systematic uncertainties into a covariance matrix. Derivatives for each systematic uncertainty are computing following
\begin{equation}
\label{eq:partial}
   \partialmusys = \vec{\mu}_{\rm nom} - \vec{\mu}_{sys}
\end{equation}
where $\vec{\mu}_{\rm nom}$ are the distances from the nominal SNIa analysis and $\vec{\mu}_{sys}$ are the distances determined from a scaling of a systematic uncertainty of perturbation $\partial \rm{sys}$. The full systematic covariance matrix $C_{\rm syst}$ for all systematic sources (${\rm sys}_k$) is then built following
\begin{equation}
\label{eq:csys}
C_{_i,_j,{\rm syst}} = \sum_{k=1}^K {\frac{\partial \mu_i}{\partial {\rm sys}_k} ~ \frac{\partial \mu_j}{\partial {\rm sys}_k} ~ \sigma_k^2},
\end{equation}
which denotes the covariance between the $i^{th}$ and $j^{th}$ SNe (or redshift bin) summed over the $K$ different sources of systematic uncertainty of magnitude $\sigma_k$, where $\sigma_k$ is the best estimate for the size of that systematic uncertainty from a-priori knowledge.

\cite{Conley2010} build an $N_{\rm SNe}\times N_{\rm SNe}$ matrix accounting for the covariance of each individual SNIa, while \cite{Scolnic18} and \cite{Brout18b} build an $N_{z_{\rm bin}} \times\ N_{z_{\rm bin}}$ covariance matrix built from the redshift-binned outputs ($\langle z \rangle_{\rm bin}$, $\langle \mu \rangle_{\rm bin}$) from BBC (Eqs. \ref{eq:zz} \& \ref{eq:muz}) instead of the individual values ($z_i$, $\mu_i$).

The systematic covariance matrix $C_{\rm syst}$ is combined with statistical uncertainties that form the diagonal matrix $C_{\rm stat}$. In the unbinned approach, $C_{\rm stat}$ is the diagonal matrix of distance uncertainties $\sigma_\mu$; in the redshift binned approach, $C_{\rm stat}$ is defined as the BBC-fitted inverse-variance weighted $\mu$-uncertainty of each redshift bin. The final covariance matrix 
\begin{equation}
\label{eq:cstatplussyst}
C_{{\rm stat+syst}} = C_{\rm stat} + C_{{\rm syst}}
\end{equation}
and is used to constrain cosmological models following
\begin{equation}
\label{eq:covmatchisq}
\chi_C^2 = \vec{D}^T~C_{\rm stat+syst}^{-1}~\vec{D}
\end{equation}
\begin{equation*}
\vec{D}= \vec{\mu}_{\rm nom} - \vec{\mu}_{{\rm mod}}
\end{equation*}
where in the redshift-binned case, $\vec{\mu}$ is computed from the binned output of BBC ($\langle \mu \rangle_{\rm bin}$) and the model distances $\mu_{{\rm mod}}$ (Eq.~\ref{eq:modeldist}) for each of the redshift bins $\langle z \rangle_{\rm bin}$.

For the covariance-matrix approach described here, we utilize \texttt{CosmoMC} \citep{cosmomc} to minimize $\chi_C^2$ to provide constraints on cosmological parameters. We include a prior on $\Omega_M$ of $0.3\pm0.01$.

\begin{figure*}[]
\centering
\includegraphics[width=.32\textwidth]{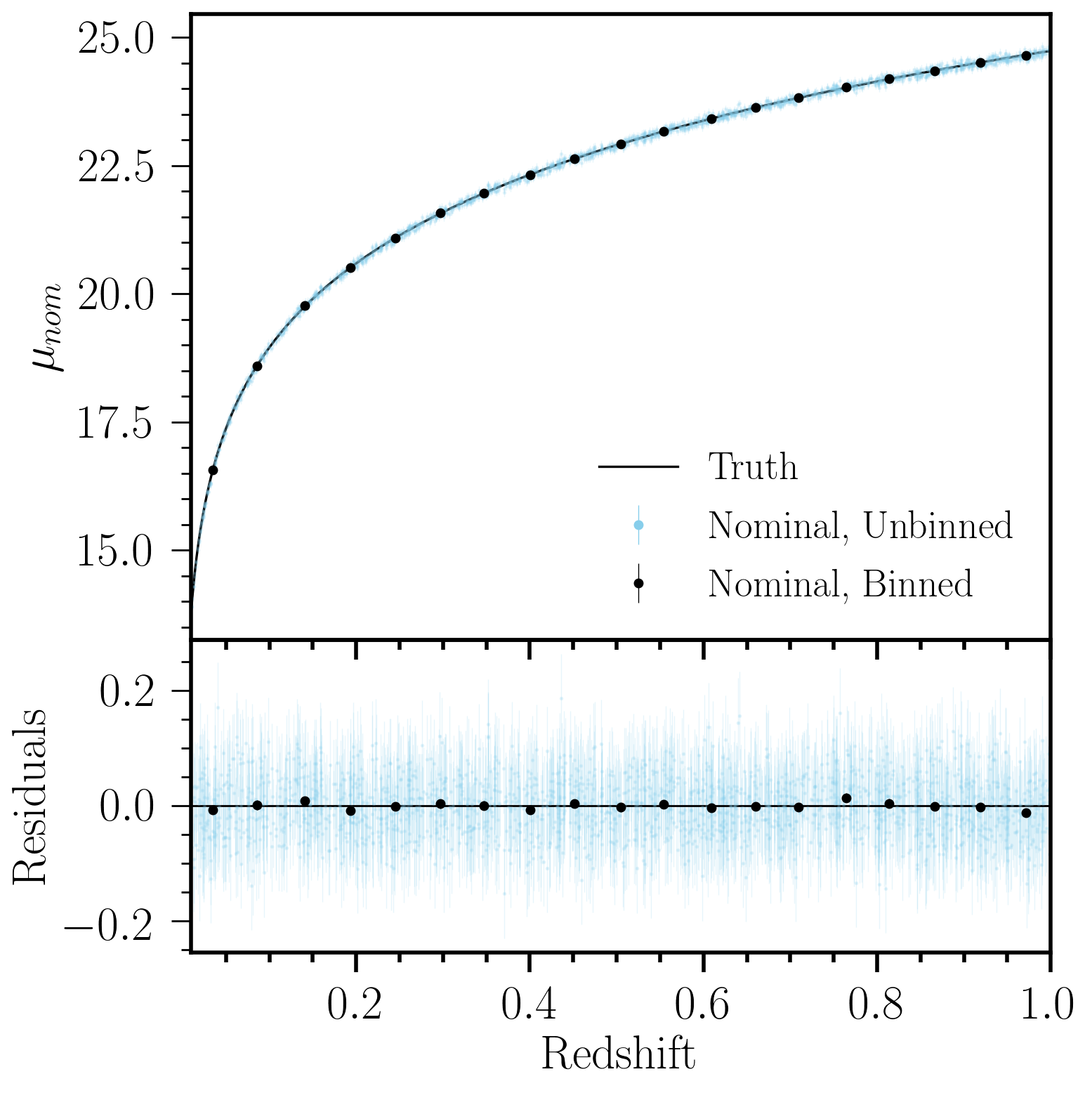}
\includegraphics[width=.32\textwidth]{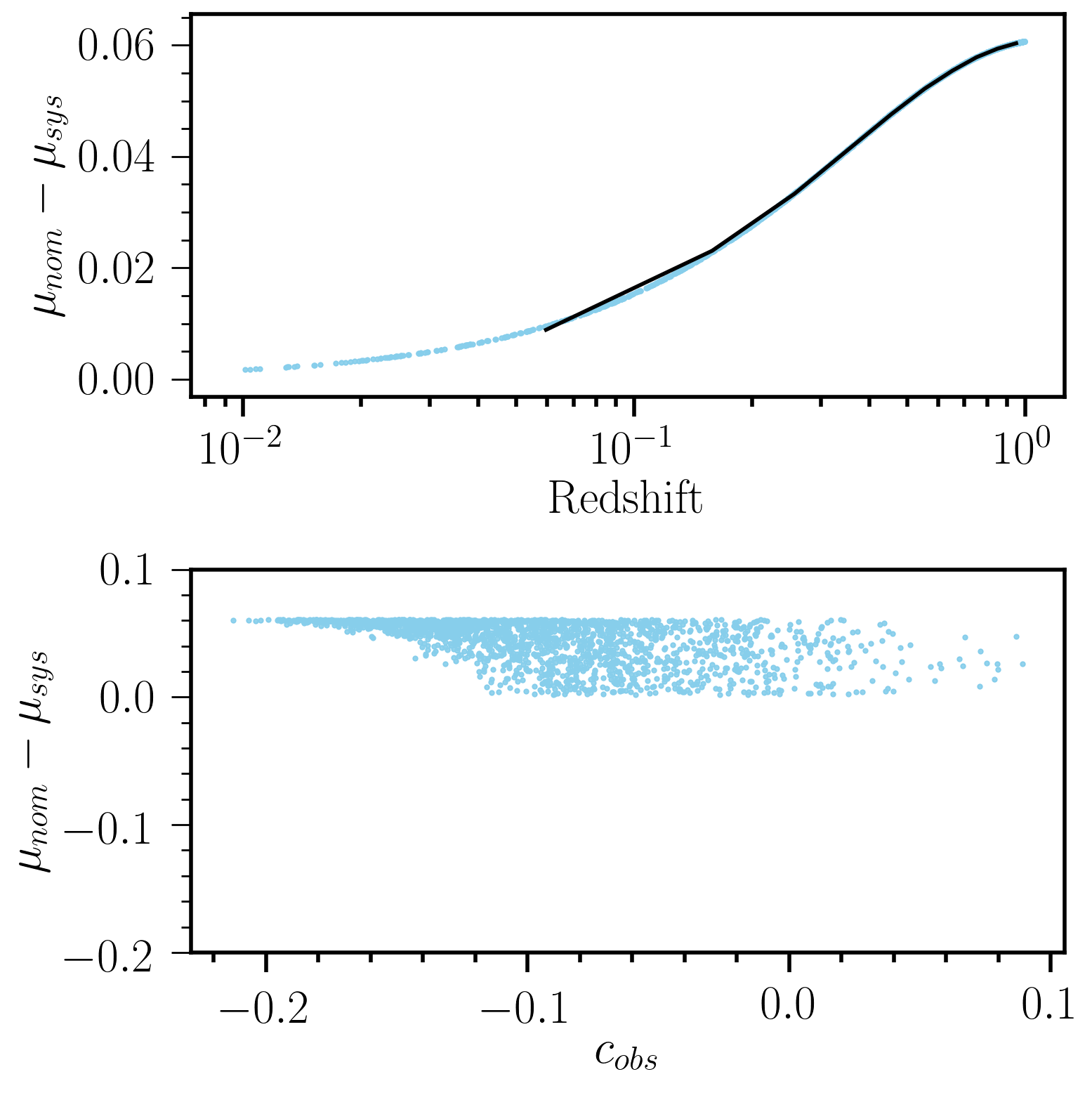}
\includegraphics[width=.32\textwidth]{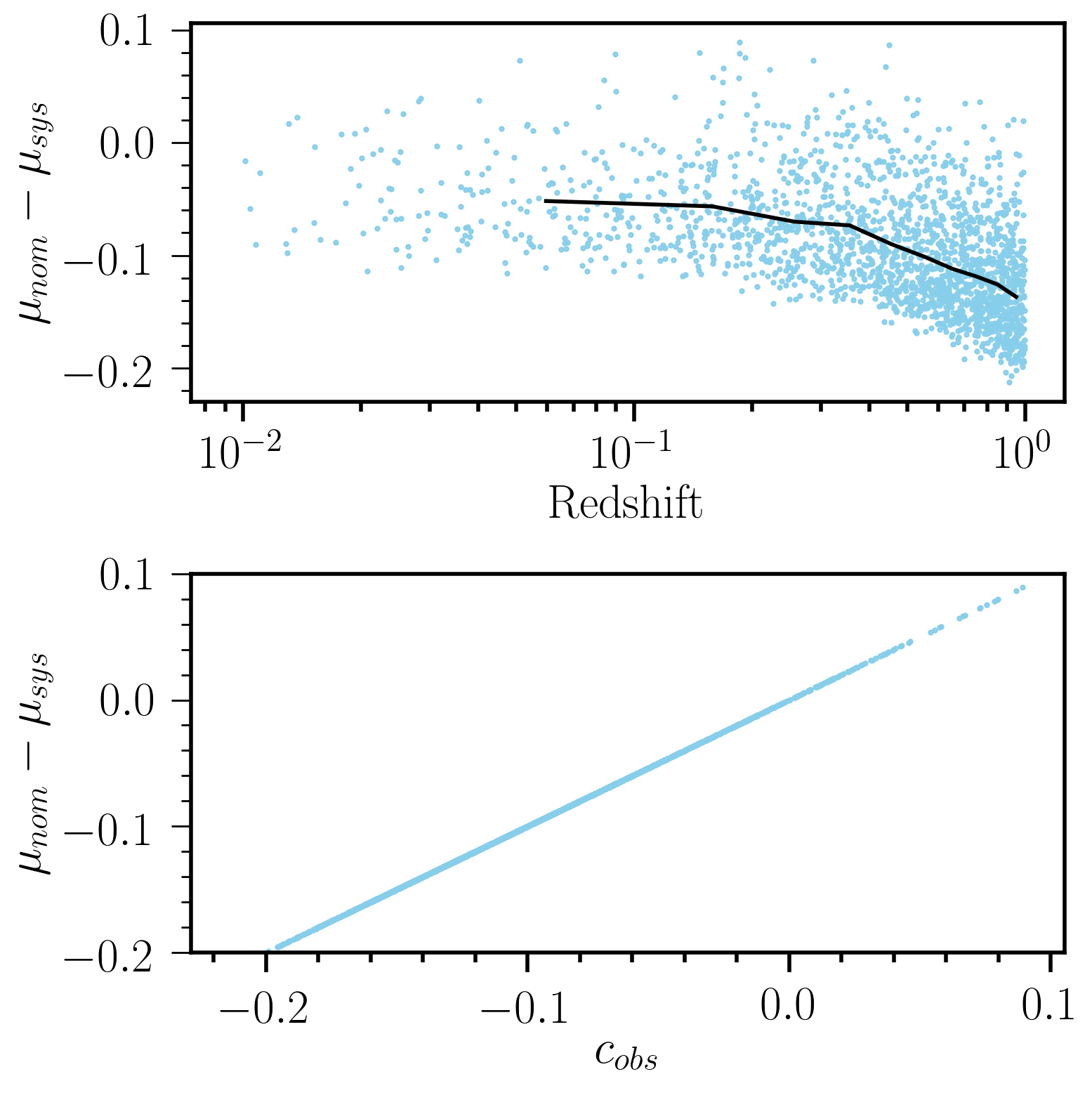}
\caption{\textbf{(Left)} Hubble diagram (top) and distance modulus residuals (bottom) of toy-model simulation, showing binned distance modulus values overlaid on top of the unbinned Hubble diagram. \textbf{(Middle)} Toy redshift-dependent systematic. The top panel shows the change in distance modulus versus redshift; the bottom panel shows the change in distance modulus versus color. Redshift-binned dataset is overlaid in black. \textbf{(Right)} Similar to the middle panel, but instead for the color-dependent systematic. \label{fig:toy_hubble}}
\end{figure*}

\subsection{Self-Calibrating Systematics Without Covariance Matrices}
\label{sec:selfcal}
We employ a secondary approach to the determination of cosmological parameters while accounting for systematic uncertainties. This approach was introduced by F11 in which cosmological parameters are determined simultaneously with the `self-calibrating' size ($S_{sys}$) of systematic uncertainties and does not involve the use of covariance matrices. Here we define $\Delta$ as
\begin{equation}
\label{eq:delta}
\vec{\Delta} = \vec{\mu}_{\rm nom} + S_{sys} \times \partialmusys - \vec{\mu}_{\rm mod}(\Omega_M,w,\rm{H}_0)
\end{equation}
where the systematic derivative ($\partialmusys$) is the same as that defined in Eq.~\ref{eq:partial}.  The following $\chi_S^2$ is minimized and facilitates constraints on $S_{sys}$ jointly with cosmological parameters ($\Omega_M,w,\rm{H}_0$):
\begin{equation}
\label{eq:scalechisq}
\chi_S^2 = \sum\vec{\Delta}^2/~\vec{\sigma}^2_\mu~
\end{equation}
where only statistical distance modulus uncertainties ($\vec{\sigma}_\mu$) are used. Similarly to Section \ref{sec:covmats}, constraints can also be computed with redshift-binned Hubble diagrams. In this case, Eq.~\ref{eq:delta} is instead computed with the binned-distance BBC outputs $\langle \mu_{\rm nom} \rangle_{\rm bin}$ and corresponding binned-model distances $\mu_{{\rm mod}}$.

We minimize $\chi_S^2$ with the \texttt{emcee} \citep{emcee} and apply a Gaussian prior (of width $\sigma_k$) on $S_{\rm sys}$, which is the same as the best estimate of the systematic size as that used in the covariance matrix approach of Eq.~\ref{eq:csys}. Again, we include a prior on $\Omega_M$ of $0.3\pm0.01$.

\section{Toy Model Showing The Impact of Binning on Systematic Uncertainties}
\label{sec:toymodel}

In this section, we provide a self-contained toy SNIa cosmological model that allows for a direct comparison between binned and unbinned analyses.  To increase accessibility, the code, plots, and MCMC chains are openly available at \href{https://github.com/Samreay/BinningIsSinning}{https://github.com/Samreay/BinningIsSinning}.\\

\subsection{Toy Model Construction}

For our toy model, we assume a simplified version of Eq.~\ref{eq:tripp} where the standardized brightness $\mu$ is defined only with a brightness term and a color term. We simulate the peak brightnesses of SNe following 
\begin{align}
m_B=\mu_{\rm mod} + \beta_{\rm sim} c + M_B  ,
\end{align}
where $\mu_{\rm mod}$ depends on the redshift and the cosmological model, $c$ is the SNIa color and is correlated to the luminosity by $\beta_{\rm sim}$, and $M_B$ is the absolute brightness of a SNIa. For model simplicity we do not include a secondary luminosity-stretch correlation as a single color-luminosity relation is sufficient to demonstrate the impact of binning.  

We then recover a distance modulus from the measurements of $m_B$ and $c$
\begin{align}
\mu_{\rm meas}= m_b + \beta_{\rm meas} c - M_B  ,
\end{align}
where here we have assumed a $\beta_{\rm meas}$ that need not equal $\beta_{\rm sim}$ as we will vary this as a systematic.

The toy model of 2000 SNIa is constructed as follows:
\begin{enumerate}
    \item Sample redshifts ($z$) from $\mathcal{U}(0.01, 1.0)$.
    \item Sample $M_B$ from $\mathcal{N}(\mu=-19.36,\sigma=0.05)$ to provide intrinsic dispersion with $\sigma_{\rm int} = 0.05$.
    \item Sample $c$ from $\mathcal{N_{\rm skew}}(\mu=-0.15, \sigma=0.07,\alpha=7)$.
    \item Add an observed color-redshift dependence as $c \mathrel{+}= -0.1z + \langle 0.1\vec{z}\rangle$, with $\langle0.1\vec{z}\rangle$ computed over all redshifts ($\vec{z}$) to roughly match that observed by JLA and Pantheon.
    \item Calculate $m_B$ using $\mu_{\rm mod} + M_B + \beta_{\rm sim} c = m_B$, with $\beta_{\rm sim}=3.1$ and $\mu_{\rm mod}$ calculated from a fiducial Flat $w$CDM cosmology with $\Omega_m=0.3$, $w=-1$.
    \item Set uncertainties $\sigma_{m_B} = 0.05$ and $\sigma_c = 0.01$, without perturbing the values drawn previously, such that we have perfect observation, but still report a value uncertainty. We confirm these results are consistent if we do add extra normal scatter corresponding to the reported uncertainty.
    \item The sample is assumed to be fit with some $\beta_{\rm meas}$ where for the nominal analysis of the toy model we assume that $\beta_{\rm meas}=\beta_{\rm sim}=3.1$.
\end{enumerate}
We do not include any additional observational or astrophysical effects, bias corrections, or any other systematic effects. The uncertainty on the distance modulus value includes contributions from $m_B$, $c$ and $\sigma_{\rm int}$,
\begin{align}
    \sigma_{\mu_{\rm nom}}^2 = \sigma_{m_B}^2 + \beta_{\rm meas}^2 \sigma_c^2 + \sigma_{\rm int}^2 .
\end{align}

Our toy sample of unbinned events is shown in the left-hand panel of Fig.~\ref{fig:toy_hubble}; we also bin the events into uniformly spaced redshift bins following Eqs. \ref{eq:zz} \& \ref{eq:muz}.

\begin{figure}
\centering
\includegraphics[width=.9\columnwidth]{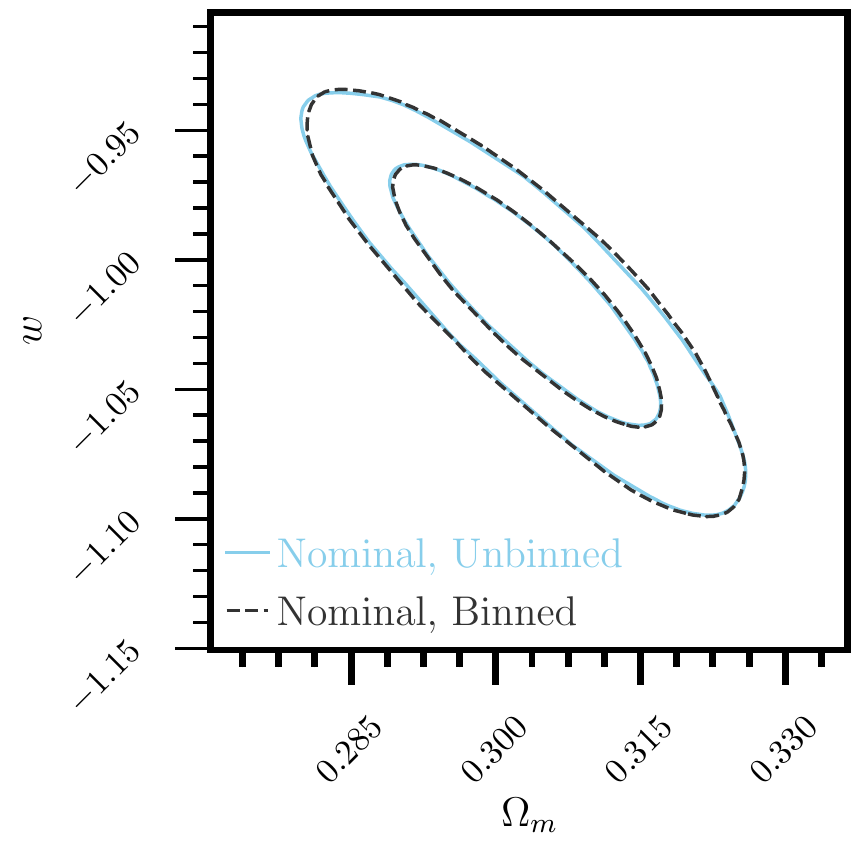}
\caption{
Comparison of the cosmological constraints on $w$ versus $\Omega_M$ for the statistical-only sample using the binned or unbinned versions of the toy dataset Hubble diagram.  In the case of statistical-only constraints, the covariance-matrix approach and scaling approach are mathematically identical (Eqs.~\ref{eq:covmatchisq} \& \ref{eq:scalechisq}), thus we only show one set of contours.}
\label{fig:baseline}
\end{figure}

The toy data ($\mu_{\rm nom}$) and uncertainties ($\sigma_{\mu_{\rm nom}}$) are used to fit a cosmological model. In Figure~\ref{fig:baseline}, we show for the statistical-only fit, constraints on cosmological parameters agree between the binned and unbinned cases.

\subsection{Toy Model Systematics}

We now study the impact of two systematics in the analysis which we refer to as `toy' systematics. First, we investigate a redshift-dependent shift in $m_B$. We compute $\partialmusys$ of Eq.~\ref{eq:partial} for this systematic by applying the following cosmology-dependent offset:
\begin{align}
    \partialmusys =  {\vec\mu}_{\rm mod}(z)|_{w=-1} - \vec{\mu}_{\rm mod}(z)|_{w=-1.15}
    \label{sysz}
\end{align}
For a second systematic, we allow for a color-dependent systematic, such that 
\begin{align}
    \partialmusys = -c~.
    \label{sysc}
\end{align}
which corresponds to an incorrect $\beta_{\rm meas}$, specifically $\beta_{\rm meas}-\beta_{\rm sim}=\Delta\beta=1$. Both the redshift-dependent (middle panel) and color-dependent (right panel) systematics are shown in Figure~\ref{fig:datatoy}.

Covariance matrices for each of these systematics are built separately following Eq.~\ref{eq:csys} where $K=1$ as we are choosing to investigate these systematics and minimize $\chi_C^2$ independently. Likewise, we also perform the `self-calibrating' approach of Section \ref{sec:selfcal} and minimize $\chi^2_S$ of Eq.~\ref{eq:scalechisq} separately for each of our two systematics.

\subsection{Toy Model Results}

\begin{figure*}[]
\centering
\includegraphics[width=0.48\textwidth]{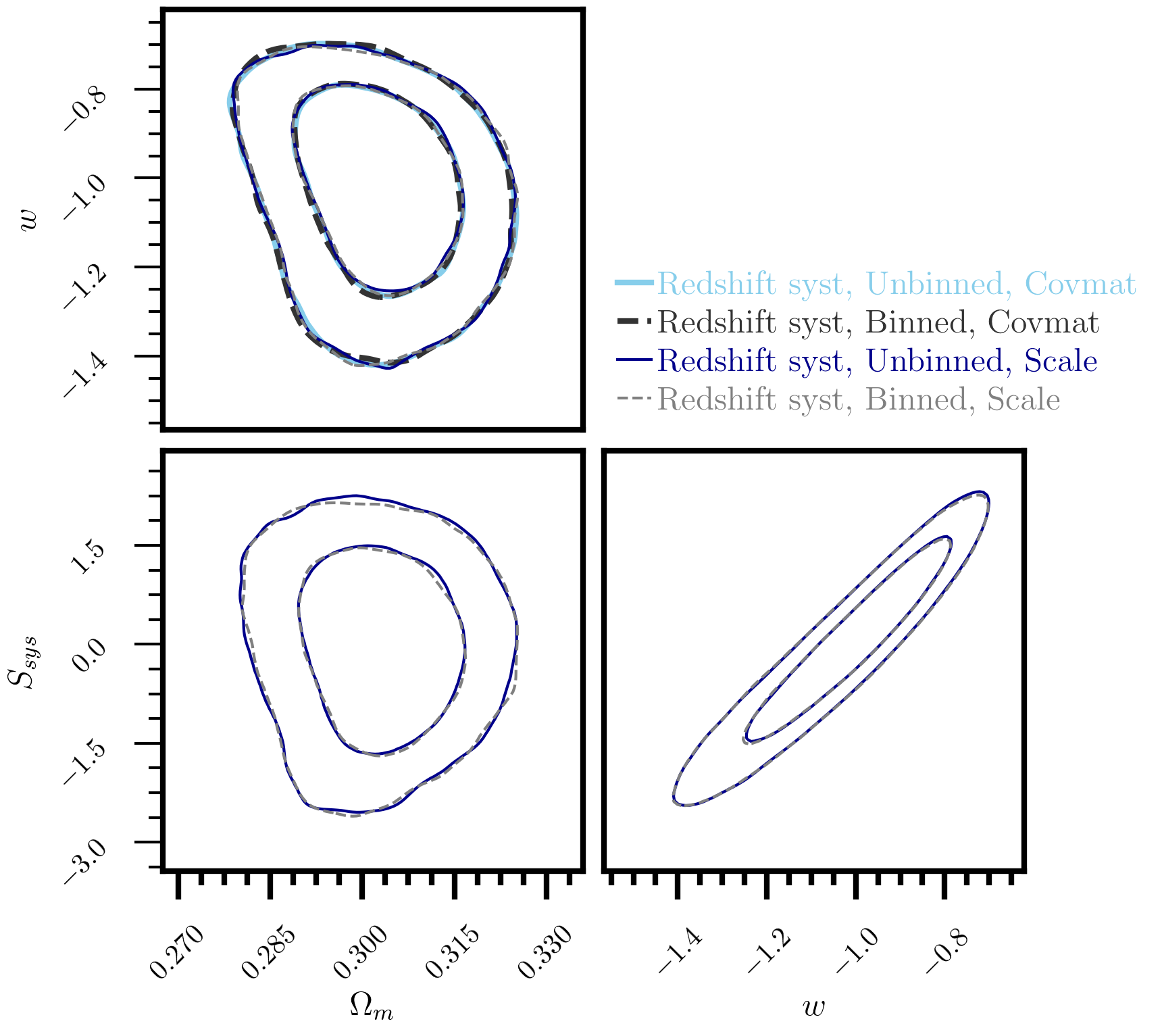}
\includegraphics[width=0.48\textwidth]{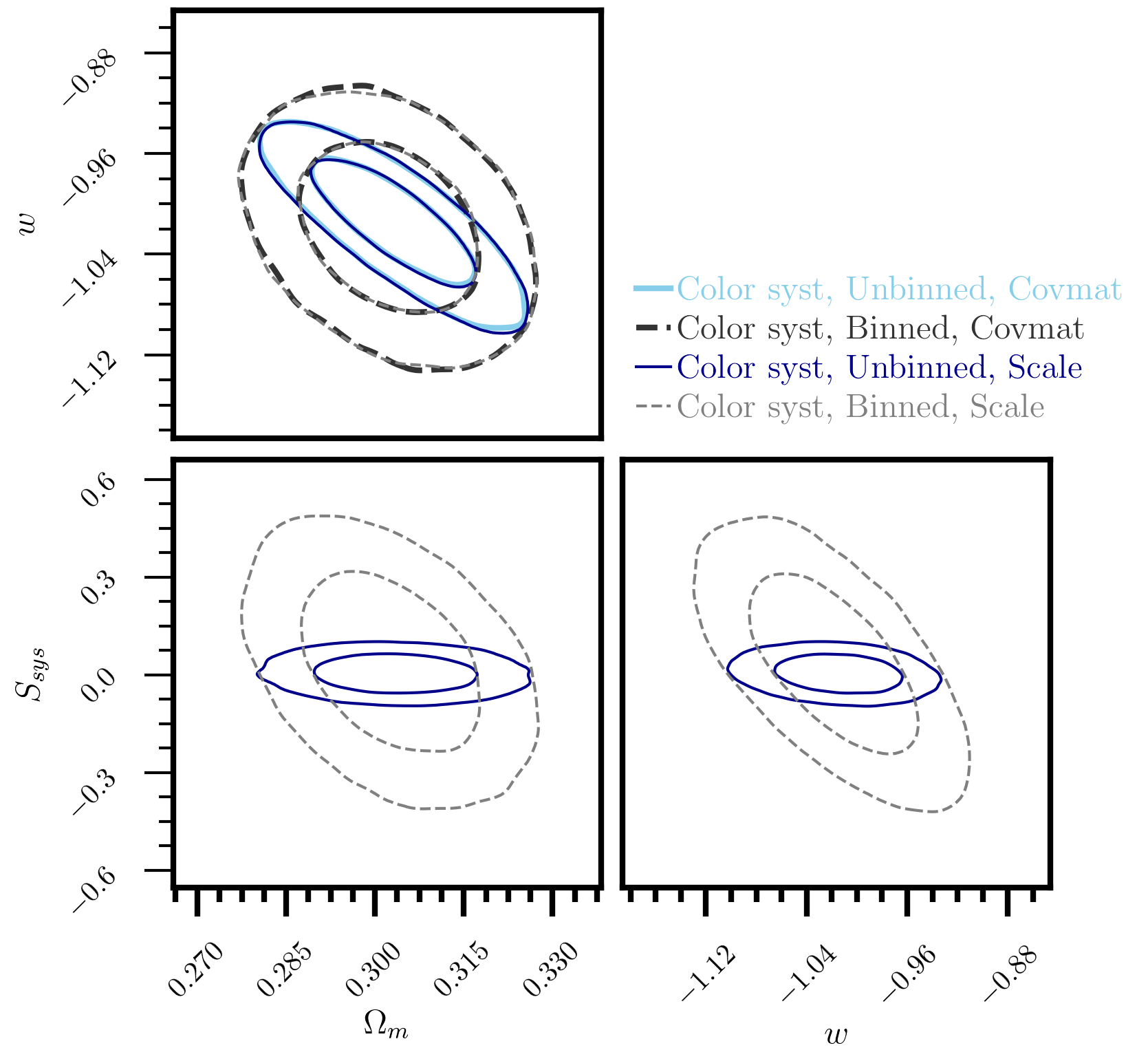}
\caption{Cosmological constraints on the toy-model simulations for the `toy' systematic uncertainties. Covariance-matrix (`Covmat') and systematic-scaling (`Scale') approaches are overlaid.  \textbf{(Left)} Constraints when including the redshift-dependent systematic for the binned (dashed) and unbinned approaches (solid). The contours are equivalent within statistical precision.  \textbf{(Right)} Constraints when including the color-dependent systematic for the binned (dashed) and unbinned (solid) versions.  The constraint on $w$ and $\Omega_M$ is $\sim2\times$ smaller when using the unbinned dataset.  }
\label{fig:toy_fits}
\end{figure*}

\begin{figure*}[]
\centering
\includegraphics[width=0.42\textwidth]{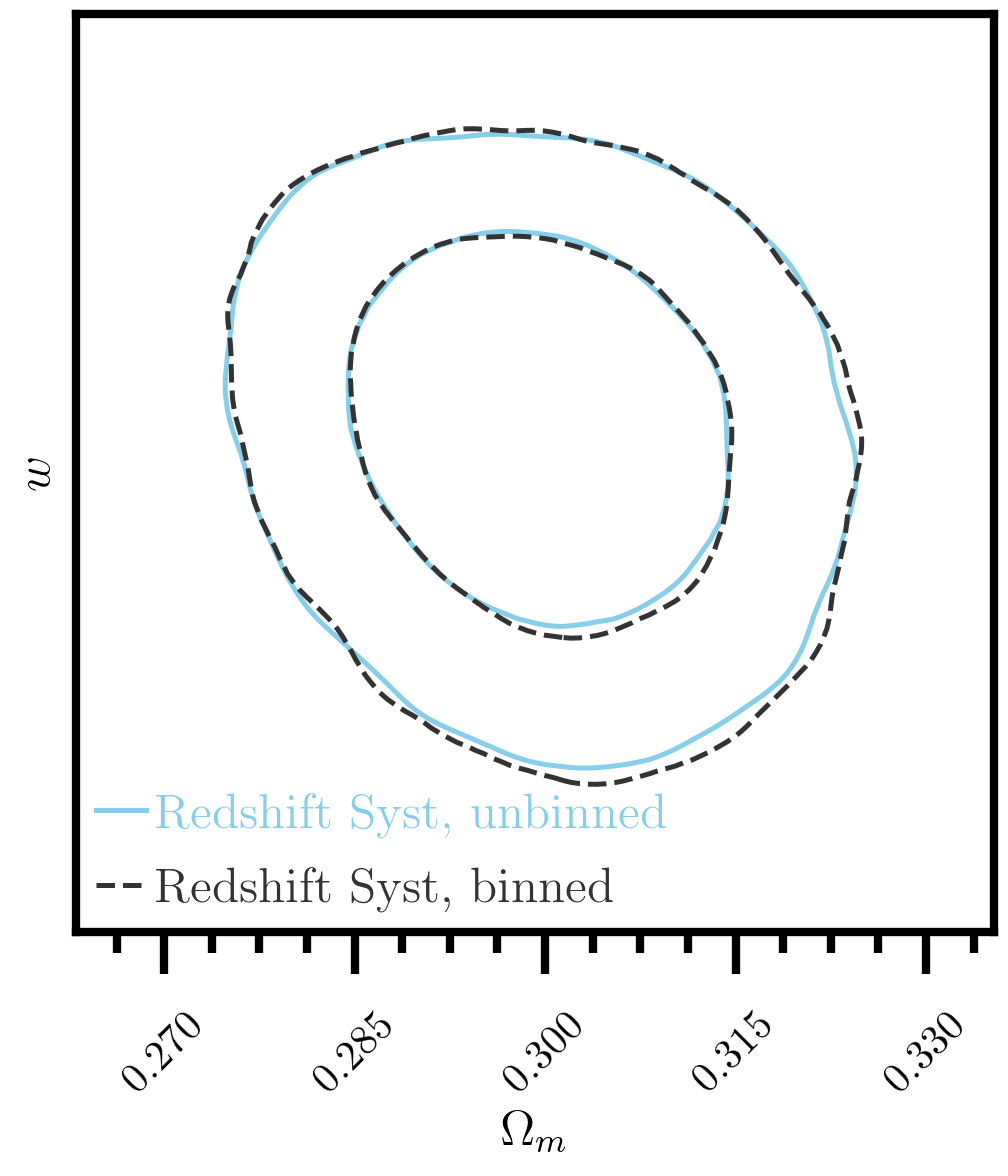}
\includegraphics[width=0.42\textwidth]{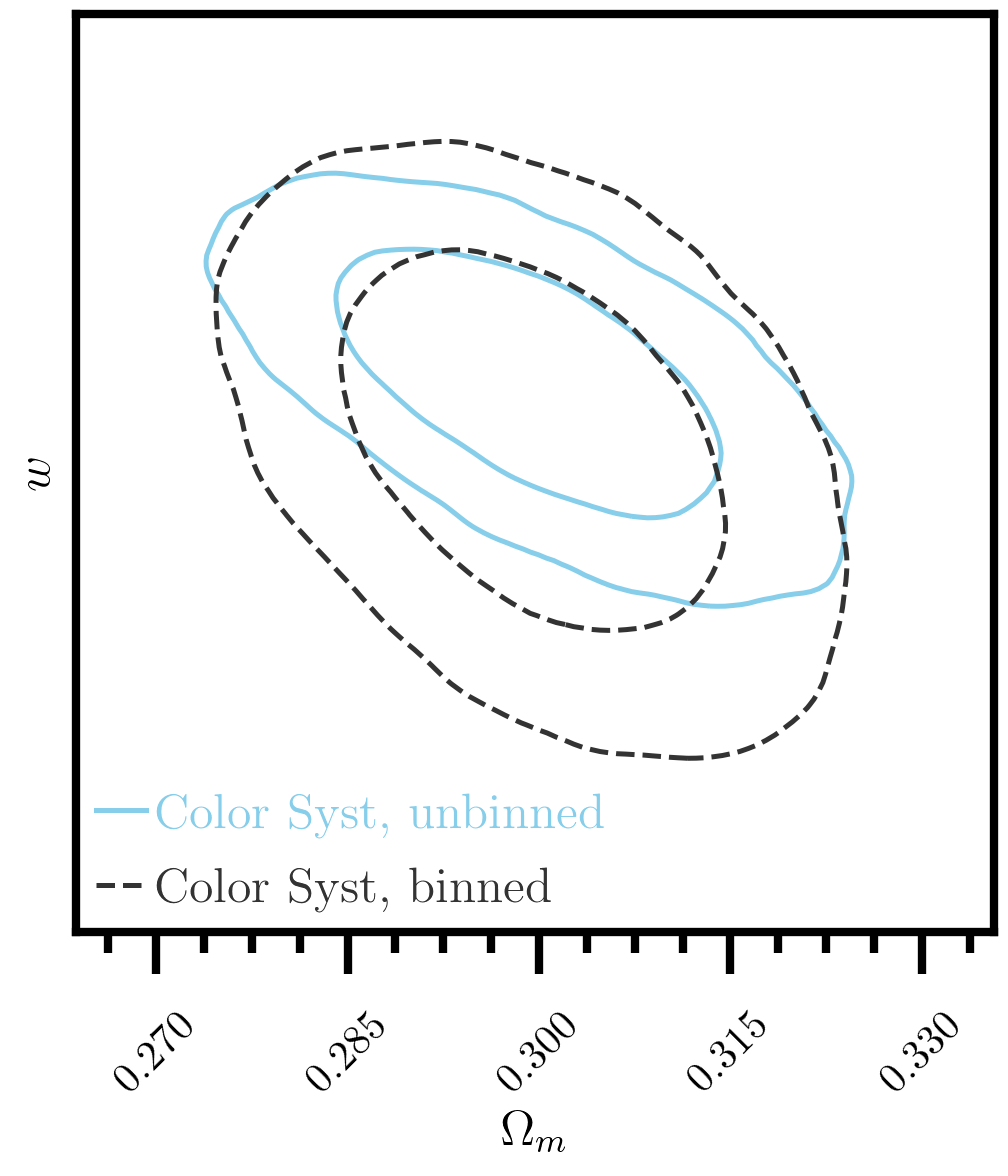}
\caption{
Cosmological constraints with real data using the covariance matrices built for the `toy' systematics. $w$ is blinded. \textbf{(Left)} Systematics that are predominantly redshift-dependent do not exhibit differences between the binned (dashed) and unbinned (solid) approaches. \textbf{(Right)} Systematics that are predominantly color-dependent do exhibit differences.  \label{fig:datatoy}}
\end{figure*}

For our two potential systematics, and our decision to bin or not to bin the dataset, we demonstrate fits using both the covariance matrix approach and the self-calibrating scale approach in Figure~\ref{fig:toy_fits} and Table~\ref{tab:toytable}. Agreement between binned and unbinned approaches is only found for the redshift-dependent systematic (Eq.~\ref{sysz}). However, there is a stark difference between binned and unbinned constraints for the color-dependent systematic (Eq.~\ref{sysc}). In both cases constraints on cosmological parameters agree between the scaling and covariance approaches, but the utility of the scaling approach is in the intuition it provides.  It clearly demonstrates that when binning the datasets, the color-dependent systematic scale $S_{sys}$ becomes degenerate with cosmological parameters, whereas for the unbinned case where all information is preserved, $S_{sys}$ can be self-calibrated to near-zero. Furthermore, the agreement between the scaling approach and covariance matrix approach for the color-dependent systematic suggests that the covariance matrix has the identical ability to `self-calibrate' the size of the systematic uncertainty despite lacking an explicit scaling parameter.

For the numerical results summarized in the top of Table~\ref{tab:toytable},
the systematic component of the $w$-uncertainty is defined as:
\begin{align}
    \sigma_w^{sys} = \sqrt{\sigma^2_{w_{stat+syst}}-\sigma^2_{w_{stat}}}~,
\end{align}
where $\sigma_{w_{stat+syst}}$ is the marginalized posterior uncertainty from the cosmological constraints for each systematic and $\sigma_{w_{stat}}$ is the uncertainty from statistical-only constraints with $\partialmusys = 0$.
We find that uncertainties on $w$ for the binned and unbinned approaches agree on the redshift-dependent systematic ($\sim0.151$). This is expected as a redshift binned dataset does not result in a loss of information for a redshift-dependent systematic. However, $w$-uncertainties on the color-dependent systematic are much smaller for the unbinned dataset ($\sim0.001$) in comparison to the binned dataset ($\sim0.050$). Furthermore, we note that the $w$-uncertainty for the unbinned case is not sensitive to the size of $\partialmusys$.

In all cases, $S_{sys}$ is consistent with zero because while we include a systematic of $\Delta\beta=1$, we have used the `correct' $\beta_{\rm meas}$ in the nominal distances ($\mu_{\rm nom}$). What is particularly informative is the uncertainty on the systematic scale ($\sigma_S$) which for the color-dependent systematic is well constrained for the unbinned dataset (0.04), but is not constrained for the binned dataset (0.18). When the systematic scale is degenerate with cosmological parameters in the fit (larger $\sigma_S$), the marginalized cosmological parameter posteriors degrade. If the systematic is fully degenerate with the cosmological model, the constraint on $S_{sys}$ comes from the prior. In either case, this results in systematics of equal or smaller size than the a-priori knowledge expressed in $\sigma_k$.

\section{Data and Realistic Simulations}
\label{sec:datasim}

In this section, we produce binned and unbinned cosmological constraints with more realistic and detailed survey simulations as well as with real SNeIa from recently published data. Despite the added complexity, the conclusion of the following section remains the same as that for the toy simulations. We also show the improvements to full Pantheon-like $w$-error budgets.

\subsection{Data}
We use a compilation of publicly available, spectroscopically classified, photometric light-curves of SNeIa that have been used in past cosmological analyses and that have been calibrated to the SuperCal system \citep{Scolnic2015}. The low-redshift (low-$z$) SNe used here are made up of, in part, by those used in B19b which are from CSP \citep{Stritzinger2010} and CfA3-4 \citep{Hicken09b,Hicken09a,Hicken12}. At low-$z$, we also include the recently released 180 low-$z$ SNe from the Foundation sample \citep{Foley17}. At high-z, we include SNeIa from PS1 \citep{Rest14,Scolnic18}, SDSS \citep{Sako11} and SNLS \citep{Betoule2014} as was done in the Pantheon analysis. Finally, we include data from the recently released DES 3-year sample \citep{Brout18b,Brout18a}. This is the same compilation of SNeIa that was used in \cite{bs20}. The cumulative redshift distribution and Hubble diagram residuals can be seen in Fig.~1 of \cite{bs20}.

This analysis uses the host-galaxy mass estimates provided by past analyses. We adopt the same masses released in the Pantheon sample, and references therein, for SDSS, PS1, SNLS, CSPDR2, and CfA. For DES3YR masses, we use the updated masses provided by \cite{Smith2020,Wiseman2020}. For the Foundation sample, we utilize masses derived in \cite{Jones18}.

 Typical selection cuts are applied on the observed data sample as was done in \cite{bs20}: we require fitted color uncertainty $< 0.05$, fitted stretch uncertainty $< 1$, fitted light-curve peak date uncertainty $< 2$, light-curve fit probability (from \texttt{SNANA}) $> 0.01$, and Chauvenaut's criterion is applied to distance modulus residuals, relative to the best fit cosmological model, at 3.5$\sigma$. In total, after selection cuts, there are 1445 SNe in this sample.

\subsection{Realistic Simulations}

We also create realistic simulations of our spectroscopically confirmed data samples. Following \cite{bs20}, we utilize the \texttt{SNANA} framework; an overview of \texttt{SNANA} simulations is given in \cite{Kessler18}.  Details of the simulations of the surveys compiled here are provided in \cite{bs20}. In short, we assume the intrinsic scatter model from \cite{Guy2010}, light-curve parameter parent populations from \cite{Scolnic16}, and survey specific simulation details for PS1 and low-$z$ samples are described in \cite{Scolnic18}; for DES, \cite{Brout18b}; for Foundation, \cite{Jones18}; and the SDSS and SNLS specifics can be found in \cite{Kessler13}. 

We produce 10 realizations of dataset-sized simulations ($\sim1500$ SNeIa). Unlike the toy model, these simulations contain selection effects, astrophysical and observational effects, cosmological effects, and we apply bias corrections to correct for the expected biases  arising from selection effects following \cite{BBC}. These simulations do not, however, have any input sources of systematic uncertainty. So similarly to what was found in the top of Table \ref{tab:toytable} for $S_{sys}$, we expect that investigated sources of systematic uncertainty will result in scales ($S_{sys}$) centered at zero.

\subsection{Results for Data and Realistic Simulations}

\begin{table}
\caption{Results for Fake Systematics}
\makebox[.4\textwidth]{%
\resizebox{.55\textwidth}{!}{

\begin{tabular}{llllrrrr}
\toprule
Data &   Binning & Method & Systematic &  \footnote{Uncertainty on simulation is 1$\sigma$ deviation in the 10 realizations.}$\sigma^{sys}_w$\ &  $\Delta_w$\footnote{$\Delta_w=w_{\rm stat}-w_{\rm sys}$} & $S_{sys}$ & $\sigma_S$\\
\hline

      \\
       Toy &    Both & - & Stat Only &    0.000 &                           0.000 &-&- \\ \\
       
       Toy &    Binned & Covmat&   $z$ Dep. &                0.151 &   -0.032 &-&- \\
       Toy &    Binned & Scaling&   $z$ Dep. &               0.151 &   -0.032 & 0.12 & 1.03 \\
       Toy &  Unbinned & Covmat&   $z$ Dep. &                0.151 &   -0.012 &-&- \\
       Toy &  Unbinned & Scaling&   $z$ Dep. &               0.151 &   -0.012 & 0.00 & 1.00 \\ \\
       
       Toy &    Binned & Covmat&   $c$ Dep. &                0.050 &    -0.026 &-&- \\
       Toy &    Binned & Scaling&   $c$ Dep. &               0.052 &    -0.028 & 0.03 & 0.18 \\
       Toy &  Unbinned & Covmat&   $c$ Dep. &                0.001 &    -0.013 &-&- \\ 
       Toy &  Unbinned & Scaling&   $c$ Dep. &               0.001 &    -0.013 & 0.01 & 0.04 \\ \\
    \hline
    \\

       10 Sims &    Both & - & Stat Only &    0.000 &                           0.000 &-&- \\ \\
       
       10 Sims &    Binned & Covmat&   $z$ Dep. &                $0.145 \pm 0.023$ &   $-0.016 \pm 0.018$ &-&- \\
       10 Sims &    Binned & Scaling&   $z$ Dep. &               $0.147\pm0.019$ &  $-0.003\pm0.042$  & $0.01\pm0.30$ & $1.03\pm0.14$ \\
       10 Sims &  Unbinned & Covmat&   $z$ Dep. &                $0.146 \pm 0.022$ &   $-0.015 \pm 0.020$ &-&- \\
       10 Sims &  Unbinned & Scaling&   $z$ Dep. &               $0.148\pm0.017$  &   $-0.013\pm0.048$ & $0.02\pm0.23$ & $1.04\pm0.12$ \\ \\
       
       10 Sims &    Binned & Covmat&   $c$ Dep. &                $0.053 \pm 0.017$ &    $-0.011 \pm 0.072$ &-&- \\
       10 Sims &  Binned & Scaling&   $c$ Dep. &               $0.055 \pm 0.018$ &  $-0.006 \pm 0.079$   & $0.02 \pm 0.35$ & $0.23 \pm 0.10$ \\
       10 Sims &  Unbinned & Covmat&   $c$ Dep. &                $0.009 \pm 0.002$ &   $-0.014 \pm 0.009$ &-&- \\ 
       10 Sims &    Unbinned & Scaling&   $c$ Dep. &               $0.010 \pm 0.002$ & $-0.010 \pm 0.010$ & $-0.06 \pm 0.04$ & $0.04 \pm 0.01$ \\ \\

    \hline
    \\
    
       Real &    Both & - & Stat Only &    0.000 &                           0.000 &-&- \\ \\
       
       Real &    Binned & Covmat&   $z$ Dep. &                0.144 &  -0.019 &-&- \\
       Real &    Binned & Scaling&   $z$ Dep. &               0.144 &   -0.028 & 0.26 & 0.99 \\
       Real &  Unbinned & Covmat&   $z$ Dep. &                0.140 &   -0.013 &-&- \\
       Real &  Unbinned & Scaling&   $z$ Dep. &               0.139 &   0.002 & 0.26 & 0.96 \\ \\
       
       Real &    Binned & Covmat&   $c$ Dep. &                0.042 &    -0.022 &-&- \\
       Real &    Binned & Scaling&   $c$ Dep. &               0.042 &    -0.025 & -0.08 & 0.19 \\
       Real &  Unbinned & Covmat&   $c$ Dep. &                0.007 &   -0.004 &-&- \\ 
       Real &  Unbinned & Scaling&   $c$ Dep. &               0.004 &    -0.005 & 0.02 & 0.30 \\ \\

\hline
\label{tab:toytable}

\end{tabular}

}}
\end{table}

Following Section \ref{sec:toymodel}, we investigate the `toy' systematics that are designed to be color-dependent and redshift-dependent. The results of these tests for the real data are shown in Figure~\ref{fig:datatoy}. Results for both the real data and 10 simulation realizations are summarized in Table \ref{tab:toytable}. We find the same conclusions as those for our toy model. First, we determine that the covariance and scaling approaches produce equivalent results, which again indicates that the covariance matrix has this same ability to self-calibrate as the scaling method. Second, we find that the constraints on cosmological parameters are significantly weaker when using binned data rather than unbinned data for the color-dependent systematic, but the constraints are equivalent for the redshift-dependent systematic.

\subsection{Realistic Systematics}

\begin{table}
\caption{Results for Realistic Systematics.}
\makebox[.4\textwidth]{%
\resizebox{.55\textwidth}{!}{

\begin{tabular}{llllrr}
\toprule
Data &   Binning & Method & Systematic &  \footnote{Uncertainty on simulation is 1$\sigma$ deviation in the 10 realizations.}$\sigma^{sys}_w$ & $\Delta_w$\footnote{$\Delta_w=w_{\rm stat}-w_{\rm sys}$} \\
\hline

      \\

       10 Sims &    Both & - & Stat Only &    0.000 &                           0.000 \\ \\
       
       10 Sims &    Binned & Covmat&   All Sys. &                $0.051\pm0.004$ &  $0.019\pm0.025$ \\
       10 Sims &  Unbinned & Covmat&   All Sys. &                $0.034\pm0.002$ &   $0.038\pm0.031$ \\ \\

      10 Sims &    Binned & Covmat&   Scatter Model &                $0.018\pm0.007$ &   $0.005\pm0.023$  \\
      10 Sims &  Unbinned & Covmat&   Scatter Model &                $0.005\pm0.001$ &  $-0.001\pm0.009$  \\ \\

       10 Sims &    Binned & Covmat&   MWEBV &                $0.008\pm0.003$ &   $-0.006\pm0.008$  \\
       10 Sims &  Unbinned & Covmat&   MWEBV &                $0.005\pm0.003$ &   $-0.003\pm0.008$  \\ \\

       10 Sims &    Binned & Covmat&   SALT2 &                $0.021\pm0.004$ &   $-0.001\pm0.019$  \\
       10 Sims &  Unbinned & Covmat&   SALT2 &                $0.016\pm0.003$ &  $0.020\pm0.014$ \\ \\

       10 Sims &    Binned & Covmat&   Survey Cal. &                $0.029\pm0.003$ &   $0.019\pm0.017$  \\
       10 Sims &  Unbinned & Covmat&   Survey Cal. &                $0.022\pm0.003$ &   $0.011\pm0.026$  \\ \\
       
    \hline
    \\
       Real &    Both & - & Stat Only &    0.000 &                           0.000 \\ \\
       Real &    Binned & Covmat&   All Sys. &                0.057 &   0.081  \\
       Real &  Unbinned & Covmat&   All Sys. &                0.038 &   0.025 \\ \\

      Real &    Binned & Covmat&   Scatter Model &                0.022 &   -0.010 \\
      Real &  Unbinned & Covmat&   Scatter Model &                0.007 &   -0.008 \\ \\

       Real &    Binned & Covmat&   MWEBV &                0.016 &   0.010 \\
       Real &  Unbinned & Covmat&   MWEBV &                0.013 &   -0.006 \\ \\

       Real &    Binned & Covmat&   SALT2 &                0.041 &   0.061 \\
       Real &  Unbinned & Covmat&   SALT2 &                0.019 &   0.041 \\ \\

       Real &    Binned & Covmat&   Survey Cal. &                0.020 &   0.005 \\
       Real &  Unbinned & Covmat&   Survey Cal. &                0.022 &   0.003 \\ \\

\hline
\label{tab:realtable}
\end{tabular}

}}
\end{table}

Now that we have verified the importance in using unbinned data to allow for self-calibration of systematic uncertainties, we can quantify the relative improvements to real SNIa systematic error budgets from the transition to unbinned data instead of binned data.  As we have shown that the covariance method and scaling method are equivalent, for this exercise we only run the covariance method as it is computationally simpler.  

We investigate the following set of systematics that have been identified by recent analysis as some of the largest contributors to SNIa uncertainty budgets:
\begin{itemize}
    \item Survey Calibration: Zeropoint calibration systematics are varied for each of the filters used by different surveys that make up the dataset following \cite{Scolnic18}.
    \item Scatter Model: Two different models for intrinsic SNIa variance (\citealp{Guy2010} \& \citealp{chotard11}).
    \item SALT2: The SALT2 surfaces are varied to account for systematic uncertainty in calibration during training of the SALT2 model following \cite{Betoule2014}.
    \item MWEBV: Systematic arising from an incorrect overall normalization (4\%) of MW extinction values following \cite{Brout18b}.
\end{itemize}

In Table \ref{tab:realtable}, we present these systematics and their impact on cosmological parameter constraints for the binned and unbinned approaches on both \texttt{SNANA} simulations and the real data. The uncertainties on parameters in Table~\ref{tab:realtable} are derived from the 1$\sigma$ scatter in 10 simulations; from this we find consistency between the data and simulations. As expected, the systematics with the largest reduction in size for unbinned data relative to binned data are the ones that have the strongest dependence on some dimension that is not redshift. For example, the scatter model systematic exhibits the largest reduction in $w$-uncertainty ($\sigma_w^{sys}$), at a factor of $\sim3\times$; this is because differences between the \cite{Guy2010} and \cite{chotard11} scatter models arise from different treatments of color variations. For the other systematics, the reduction is smaller but significant. For the combination of all the systematics (All Sys.), there is a reduction in $\sigma_w^{sys}$ of $1.5\times$. We note that since the reported uncertainty on parameters in Table~\ref{tab:realtable} is the 1$\sigma$ scatter, in order to assess differences between binned and unbinned simulations for the same systematic, one should divide the reported uncertainty by $\sqrt{10}$.

\section{Discussion}
\label{sec:discussion}

In this analysis, we have shown with toy models, realistic simulations, and real data that while binning the Hubble diagram produces self-consistent statistical constraints on cosmological parameters compared to not binning, this consistency is lost when propagating systematics. This is demonstrated with two separate approaches, one utilizing a pre-computed covariance matrix, and another that includes additional parameters for `self-calibrating' scales of each systematic during the cosmological fit. We find that both approaches agree with each other when either binning or not binning, but that the unbinned data always produces equal or better constraints on cosmological parameters. This implies that the covariance matrix itself has the ability to `self-calibrate' systematics, without the need for explicitly fitting for a scale of each systematic as was done in F11. Furthermore, from our results we interpret the impact of binning as removing the self-calibration ability of the data on various systematics.  

Recent cosmological analyses have had different approaches towards the creation of systematic covariance matrices, but all marginalized over any dependence of SN systematic on parameters other than redshift.  In \cite{Betoule2014}, instead of first binning their Hubble diagram, they smooth the systematic vectors over redshift. While the exact recipe for the smoothing is not given, from their analysis it is clear this would remove any color-dependence of the SNe. Furthermore, \cite{Betoule2014} bin their Hubble diagram and covariance matrix, and show that the binned and unbinned versions produce equivalent cosmological constraints. Since this is assessed after their smoothing, we interpret this to have demonstrated that their smoothing process approximates a binning process and does not address the issue that our analysis highlights.

Other recent cosmological analyses (Pantheon, DES3YR) have used the BBC method, where systematic covariance matrices are built from the binned versions of the Hubble diagram. In \cite{Brout18b}, the binned Hubble diagram and the binned covariance matrices are used for all the cosmological fits.  In \cite{Scolnic18}, cosmological fits are done with the unbinned Hubble diagram, but the covariance matrix is built from an interpolated version of the binned systematic matrix. Therefore, for both of these analyses, we expect them to overestimate systematic uncertainties relative to the unbinned approach suggested in this paper.

Another recent cosmological analysis, \cite{Jones18}, used both BBC and PSBEAMS \citep{JonesBEAMS}, a different tool to measure a binned Hubble diagram. Crucially, this analysis is different from the aforementioned spectroscopic analyses because it was a photometrically classified sample. For photometric samples, only the binned distance estimates from BBC or PSBEAMS have a robust, frequentist meaning. This is because the datasets must be binned in order to account for and marginalize over the photometric typing probabilities and to properly account for the predicted core-collapse distributions entering as contamination in the Hubble diagram. 

Since the most widely used methods for analyzing photometrically classified samples rely on binning data, this now presents a clear challenge to address for experiments like LSST.  While for spectroscopic samples, one can continue to use BBC and build covariance matrices from the unbinned Hubble diagram, one would have to redevelop a cosmology fitter to allow for individual photometric probabilities per SN, core-collapse population simulations, and fit for cosmology simultaneously.

Alternate solutions for photometric SNIa cosmology constraints include Bayesian hierarchical models such as UNITY \citep{unity} or Steve \citep{Hinton18}. These methods have more in common with the scaling approach of Section \ref{sec:selfcal} and but also solve for the correlation coefficients in the Tripp estimator, selection effects, cosmological parameters, and systematics simultaneously. However, these methods face challenges because varying SNIa nuisance parameters causes non-linear effects that must be simulated and cannot be evaluated analytically. Another possible solution is to adopt a forward modeling approach like that of \cite{Jennings16}. Though this type of method is very computationally expensive, we argue that more effort should be put in this direction.

\begin{figure}[]
\centering
\includegraphics[width=0.45\textwidth]{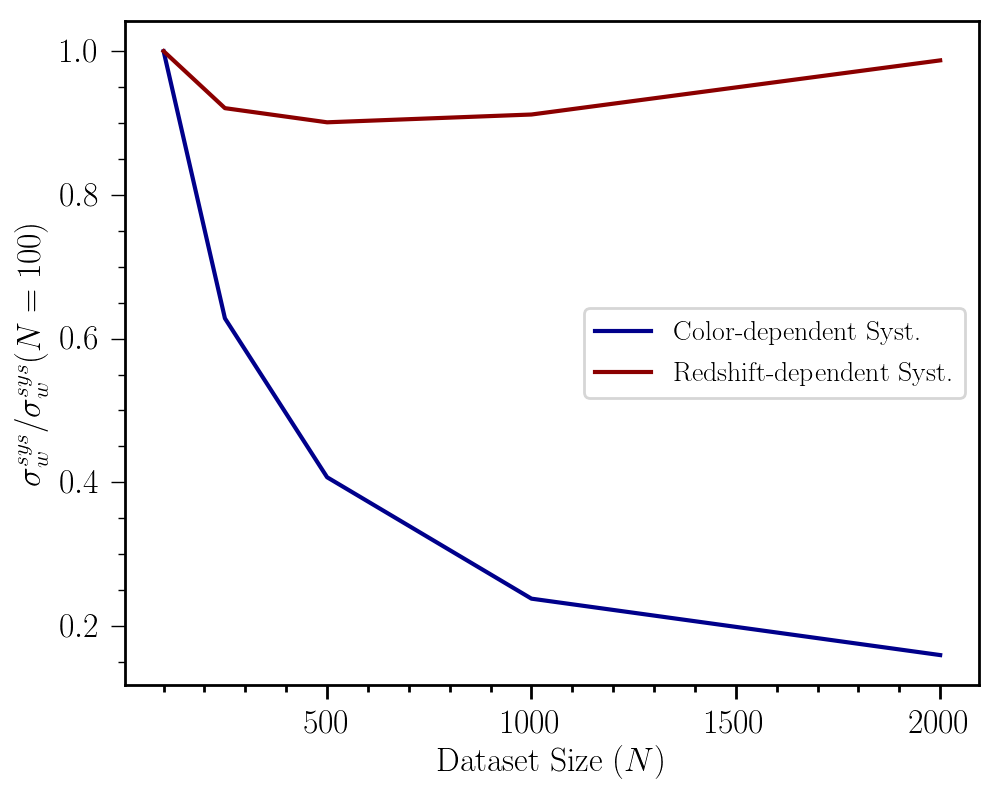}
\caption{
  \label{fig:toysizesys} Systematic floors on $w$ ($\sigma_w^{\rm sys}$) for our `toy' systematics on toy simulations of differing sizes ($N$) using the covariance-matrix approach. Improvements are normalized to the systematic uncertainty for a toy dataset of size $N=100$ SNe. Systematics that exhibit dependence other than redshift can be self-calibrated by the dataset, and as the dataset grows in size, this self-calibration ability strengthens inherently.}
\end{figure}

Whatever the path forward, it's important to forecast how self-calibration can improve systematic error budgets of future experiments like LSST or NGRST. F11 showed that future experiments will be able to self-calibrate for zeropoint uncertainties with large enough statistical samples. Here we have extended this to all sources of uncertainty and expanded beyond the F11 approach to show that the covariance matrix approach accomplishes the same task. We also found that such self-calibration of systematics is even stronger for systematics other than survey calibration (i.e. Scatter Model). Furthermore, the implications of the self-calibration of systematics by the dataset itself suggest that as the dataset size grows, the systematic floor for certain systematics will continue to shrink. This is illustrated in Fig.~\ref{fig:toysizesys} for which we have assessed the `toy' systematic uncertainties for datasets of varying sizes ($N$).

Recent forecast papers like the LSST SRD \citep{Mandelbaum18} and the NGRST forecast by \cite{Hounsell18} assessed their systematic requirements with binned, simulated datasets and covariances matrices. Thus, they likely underutilized the self-calibrating aspect of their datasets and overestimated the impact of systematic uncertainties. Despite \cite{Hounsell18} binning their dataset, they still observed some small amount self-calibration which can be seen by the plateau of the impact of certain systematics in their Figure 9, but they likely missed out on much of this self-calibrating ability. In light of our findings, we suggest both the LSST and NGRST systematic forecasts be redone. This is not a trivial suggestion as inversions of a covariance matrix with $>100,000$ SNe are computationally intensive.

While the finding in this paper implies a significant amount of work for the community to bring photometrically-classified SNIa analyses on par with spectroscopic analyses, it also means that the impact of systematics is less strong than previously thought, and that as statistics grow in future surveys, the data will be better able to self-calibrate out certain systematic uncertainties. Thus, leveraging the statistical power of the next generation of SN samples will naturally result in a lowering of the systematic floor.

\section{Conclusion}
\label{sec:conclusion}

We have shown that the method of binning the Hubble diagram in redshift space and then propagating systematic uncertainties leads to overestimates of the uncertainties of cosmological parameters.  We demonstrated using two different methods that this over-estimation is due to compression of the individual information which removes the self-calibrating power of the data to particular systematics. We find that the covariance matrix approach has the same self-calibrating power that the scale-based approach does. We show for a realistic set of systematics, that the unbinned approach results in systematic error budgets of cosmological parameters that are $\improvement\times$ lower than budgets made using the binned approach.  As the binned approach is currently a pillar of analyses of photometrically classified samples, it's imperative for the community to develop a new method that can perform photometric analyses without binning.

Finally, this paper follows along in a line of recent works \citep{kipping10,storeyfisher20,Grimmett20} on the effects of binning in astrophysical analyses. We strongly encourage further discussion and careful forethought before binning data across all areas of physics, astrophysics, and cosmology.

\bibliography{paper}
\bibliographystyle{aasjournal}
\section{Acknowledgements}
We thank Mike Jarvis for his wisdom. We thank Rick Kessler, Adam Riess, Christopher Stubbs, Alex Kim, and William Kenworthy for very useful discussions. We are appreciative of Rick Kessler for his ever-useful \texttt{\texttt{SNANA}} package. DB acknowledges support for this work was provided by NASA through the NASA Hubble Fellowship grant HST-HF2-51430.001 awarded by the Space Telescope Science Institute, which is operated by Association of Universities for Research in Astronomy, Inc., for NASA, under contract NAS5-26555. DS is supported by DOE grant DE-SC0010007 and the David and Lucile Packard Foundation. DS is supported in part by NASA under Contract No. NNG17PX03C issued through the WFIRST Science Investigation Teams Programme. Contours and parameter constraints are generated using the \textsc{ChainConsumer} package \citep{Hinton16}. Plots generated with Matplotlib \citep{matplotlib}. Usage of astropy \citep{astropy}, SciPy \citep{scipy}, and NumPy \citep{numpy}.

\end{document}